\documentstyle[12pt]{article}
\input epsf
\newcommand{\E}{{\cal E}}
\catcode`\@=11 
\newcommand{\fcaption}[1]{
        \refstepcounter{figure}
        \setbox\@tempboxa = \hbox{\rm Fig.~\thefigure. #1}
        \ifdim \wd\@tempboxa > 5.5in
           {\begin{center}
        \parbox{5.5in}{\footnotesize\baselineskip=12pt Fig.~\thefigure. #1}
            \end{center}}
        \else
             {\begin{center}
             {\tenrm Fig.~\thefigure. #1}
              \end{center}}
        \fi}
\newcommand{\lsim}{\mathrel{\mathpalette\@versim <}}
\newcommand{\gsim}{\mathrel{\mathpalette\@versim >}}
\def\@versim#1#2{\vcenter{\offinterlineskip
 \ialign{$\m@th#1\hfil##\hfil$\crcr#2\crcr\sim\crcr } }}

\catcode`\@=12 
\newcommand{\pslash}{\not{\hbox{\kern-0.1pt $p$}}}

\begin{document}

\vspace*{3.5cm}
\begin{center}
{\large\bf DESTABILIZING DIVERGENCES\\ IN SUPERGRAVITY THEORIES\\}
\vspace*{6.0ex}
{JONATHAN A. BAGGER\\[1.5ex]}
{\it  Department of Physics and Astronomy\\
Johns Hopkins University\\
Baltimore, MD 21218 \\}

\vspace*{4.5ex}
{\bf Abstract}
\end{center}
\vglue 0.3cm
{\rightskip=3pc
\leftskip=3pc
\noindent
We study the stability of the gauge hierarchy in hidden-sector
supergravity theories.  We show that a destabilizing tadpole
can appear if a theory has a gauge- and global-symmetry
singlet with renormalizable couplings to the visible fields.
We also find a quadratically divergent two-loop contribution
to the supersymmetric effective potential.  This term illustrates
the difficulty of using the ``LHC mechanism" to control quadratic
divergences in theories with Planck-scale vevs.
\vglue0.1cm}

\vspace*{3.0ex}
{\bf\noindent 1. Introduction}
\vglue 0.2cm

In a renormalizable theory, supersymmetry stabilizes the gauge
hierarchy,
\begin{equation}
M_W\ \ll\ M_P\ ,
\end{equation}
by canceling all quadratic divergences.  This cancellation
persists even if supersymmetry is explicitly broken by certain
soft operators \cite{gg}.  These two facts have sparked an
explosion of interest in the phenomenological application of
supersymmetric gauge theories.

The problem is that any realistic theory
is likely to be a {\it nonrenormalizable} effective theory, valid
below some scale $\Lambda$.  This might be the Planck scale, $M_P$,
the string scale,  $M_X$, or the unification scale, $M_G$ \cite{guts}.
For the purposes of this talk, we will take $\Lambda \simeq M_P$.

In a supersymmetric effective theory, the K\"{a}hler potential,
$K$, and the superpotential, $P$, typically contain an infinite
tower of nonrenormalizable terms, suppressed by the scale
$\Lambda$.  For the case at hand, this means
\begin{eqnarray}
K &=& \Phi^+ \Phi\ +\ \Phi^+ \Phi \left( {\Phi +
\Phi^+ \over M_P} \right)\ +\ \ldots\ \nonumber\\
P &=& \Phi^3\ +\ {1\over M_P}\, \Phi^4\ +\ \ldots
\label{nonrenorm}
\end{eqnarray}
As we shall see, these nonrenormalizable terms can reintroduce
quadratic divergences.  These divergences
have the potential to destabilize the gauge hierarchy
\cite{guts} -- \cite{jain}.

In this talk I will report on work with Erich Poppitz and Lisa Randall
\cite{bpr} in which we clarify the conditions under which radiative
corrections destabilize the gauge hierarchy.  We will find a two-loop
quadratically divergent contribution to the superspace effective potential.
We shall see how this term can destabilize the hierarchy in supersymmetric
theories with gauge- and global-symmetry singlets.  We will also discuss
how it affects the hierarchy in theories with Planck-scale vevs.

\vspace*{3.0ex}
{\bf\noindent 2. Destabilizing Divergences}
\vglue 0.2cm

Throughout this talk we will use a toy model to represent the
minimal supersymmetric standard model.   The toy model embodies
all the essential physics that we wish to discuss.  Therefore we shall
restrict our attention to a single ``Higgs" superfield, $H$, and take
the superpotential, $P$, to be
\begin{equation}
P\ =\ {1 \over 2}\ \mu\, H^2\ .
\end{equation}
With this potential, the Higgs has a mass $M_H \simeq \mu \simeq M_W$.
(A discrete $Z_2$ symmetry replaces the gauge symmetry of the standard
model.  We assume that $Z_2$ is not broken for scales larger than
$M_W$.)

The hierarchy is destabilized if radiative corrections
lift $M_H \gg M_W$.  In this model, simple superspace power counting
indicates that the hierarchy is stable \cite{old}.  With more fields,
however, the hierarchy can be destroyed.  The potentially dangerous
operators involve gauge- and global-symmetry singlet superfields.

Without loss of generality, we can distinguish two cases:

\begin{itemize}

\item Let $N$ be a gauge- and global-singlet superfield, which
couples directly to the Higgs,
\begin{equation}
P \ =\  {1 \over 2}\ \mu\, H^2\ +\ {1 \over 2}\  m\, N^2 \ +
\ \lambda\, N H^2\ +\ \ldots
\label{PN}
\end{equation}
Because of its renormalizable coupling to $H$, the field $N$ is said
to be in the visible sector.  (For the purposes of this talk, we
take $m = \mu$.)

\item Let $C$ be a gauge- and global-singlet hidden-sector superfield
{\it (un champ du secteur cach\'{e}),} which couples to the visible
sector through terms suppressed by the Planck mass, $M_P$,
\begin{equation}
P \ =\ {1 \over 2}\ \mu\, H^2\ \left[\, 1 \ + \  \left( {C \over M_P}
\right)^n \ + \ \ldots \,\right]\ .
\label{PC}
\end{equation}
\end{itemize}
Typically, the visible-sector fields $H$ and $N$ are assigned weak-scale
vevs,
\begin{eqnarray}
\langle H \rangle & \lsim &  M_W \ +\ \theta \theta M_W^2 \nonumber \\
\langle N \rangle & \lsim &  M_W \ +\ \theta \theta M_W^2\ ,
\label{hvev}
\end{eqnarray}
while the field $C$ can have a larger vev,
\begin{equation}
\langle C \rangle\ \lsim\ M_P\ +\ \theta \theta M_S^2\ ,
\label{cvev}
\end{equation}
where $M_S^2 \simeq M_W M_P$ denotes the scale of supersymmetry breaking.
The vev $\langle C \rangle$ can be fixed by the hidden-sector potential,
or it can be free, as with a string modulus, in which case we
denote $C$ by $T$.

The vevs (\ref{hvev}) and (\ref{cvev}) are typical tree-level vevs for
the fields $N$ and $C$.  They preserve the hierarchy, as can be seen
by substituting into (\ref{PN}) and (\ref{PC}).  The lowest components
of the superfield vevs give a supersymmetric renormalization of $\mu$,
while the highest components induce a supersymmetry-violating
mass for the scalar component, $h$, of the Higgs superfield, $H$.

{}From these expressions we see that the hierarchy is destabilized if
loop corrections induce vevs of the order
\begin{equation}
\langle N \rangle\ \simeq\ M_S\ +\ \theta \theta M_S^2
\end{equation}
or
\begin{equation}
\ \ \ \langle C \rangle\ \simeq\ M_P\ +\ \theta \theta M_P^2\ .
\end{equation}
These vevs are dangerous and must be avoided if supersymmetry is
to solve the gauge hierarchy problem.

\vspace*{3.0ex}
{\bf\noindent 3. Destabilizing Divergences at One Loop}

\vspace*{2.0ex}
{\it\noindent 3.1\ General Formalism}
\vglue 0.2cm

\begin{figure}[t]
\epsfysize=1.5in
\hspace*{3.0in}
\epsffile{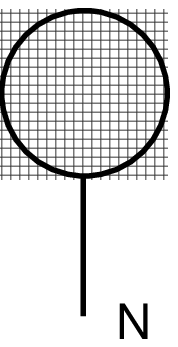}
\fcaption{A visible-sector tadpole diagram.}
\end{figure}

In flat space, the supersymmetric action is given by
\begin{equation}
S\ =\ \int d^4 \theta\  K\  +\  \left[\, \int d^2 \theta P\ +
\ h.c.\,\right]\ ,
\end{equation}
where $K$ is the K\"{a}hler potential, and $P$ is the superpotential
of the supersymmetric theory.

In curved space, the situation is more complicated.  The
superspace kinetic term is given by
\begin{equation}
- 3 M_P^2 \int d^4 \theta\ E \ e^{-K/3 M_P^2}\ ,
\label{KE}
\end{equation}
while the superspace superpotential is
\begin{equation}
\int d^2 \theta\ \E \, P\ +\ h.c.
\label{superp}
\end{equation}
In these expressions, $E$ and $\E$ are superdeterminants of the
supervielbein, superspace generalizations of the density $\det e^a_m$
familiar from general relativity.

The superspace action (\ref{KE}) plus (\ref{superp}) is manifestly
supersymmetric.  It is also invariant under super-K\"{a}hler-Weyl
transformations, because
\begin{equation}
E\ \sim\ \Sigma^+ \Sigma\ ,\qquad\qquad \E\ \sim\ \Sigma^3\ ,
\end{equation}
where $\Sigma$ is the superconformal compensator.  When supersymmetry
is broken, $\Sigma$ plays the role of a spurion,
\begin{equation}
\langle \Sigma \rangle\ \simeq \ 1\ +\ \theta \theta\, {M_S^2 \over M_P}\ .
\end{equation}
For hidden-sector scenarios, with $M^2_S \simeq M_W M_P$, $\langle \Sigma
\rangle$ contributes to the soft masses of the visible-sector particles.
(When $M_S \simeq M_W$, the vev of $\Sigma$ can be ignored.)

\vspace*{2.0ex}
{\it\noindent 3.2\ Visible-Sector Tadpoles}
\vglue 0.2cm

We are now ready to begin our analysis of destabilizing divergences in
supergravity theories.  We first consider the case of a gauge- and
global-symmetry singlet in the visible sector.  By power counting, it
is not hard to see that the dangerous diagrams are tadpoles (Fig.~1),
which scale like
\begin{eqnarray}
\delta S & \simeq & \Lambda \int d^4 \theta\ E\, N \nonumber\\
& \simeq & \Lambda \int d^4 \theta\ \Sigma^+\Sigma\, N \nonumber\\
& \simeq & M_S^2 \int d^2 \theta\ \Sigma\, N\ .
\end{eqnarray}
In hidden-sector models, such a tadpole will induce a vev of order
$M^2_S$ for the highest component of $N$.  This vev is dangerous
because it gives a mass of order $M_S$ to the Higgs scalar, $h$.

\begin{figure}[t]
\epsfysize=1.5in
\hspace*{2.8in}
\epsffile{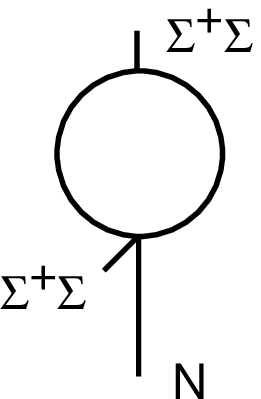}
\fcaption{In the minimal case, the one-loop quadratically divergent
contributions cancel\newline between the two $\Sigma^+\Sigma$
contributions.}
\end{figure}

\begin{figure}[t]
\epsfysize=1.5in
\hspace*{2.8in}
\epsffile{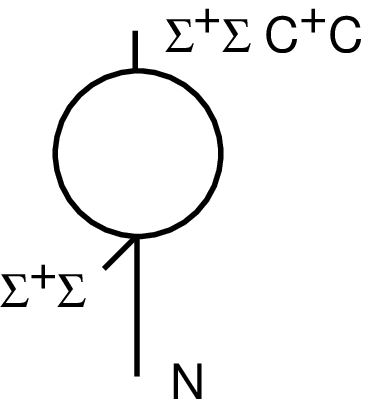}
\fcaption{In the nonminimal case, the one-loop
quadratically divergent contributions do\newline not cancel.}
\end{figure}

To determine whether such a tadpole arises, let us first consider
the minimal case, in which there is no coupling between the visible
and invisible sectors.  Therefore we take the K\"{a}hler
potential to be
\begin{equation}
K \ = \ N^+ N\ +\ C^+C\ +\ H^+ H
\left( 1\ + \ {N + N^+ \over M_P} \right) \ +\ \dots
\label{minimal}
\end{equation}
At one loop, there is one potentially dangerous superspace diagram,
as shown in Fig.~2.  Each $\Sigma^+\Sigma$ insertion induces a
quadratic divergence.  However, the two divergences exactly cancel,
so the hierarchy is stable \cite{jain}.  It is necessary to go to
two loops to see whether the cancellation is natural, or whether
it is an accident of the one-loop approximation.

Let us now consider the nonminimal case, in which there are couplings
between the visible and hidden-sector fields.  We take the
K\"{a}hler potential to be given by
\begin{equation}
K \ = \ N^+ N\ +\ C^+C\ +\ H^+ H \left( 1\ +
\ {N + N^+ \over M_P}\ +\ { C C^+ \over M^2_P} \right) \ +\ \ldots
\label{nonminimal}
\end{equation}
As before, there is one potentially dangerous diagram,
as shown in Fig.~3.  Now, however, when $C$ gets a vev,
the extra term in $K$ spoils the cancellation between the two
quadratic divergences.  A one-loop tadpole is induced
\cite{old,jain}
\begin{eqnarray}
\delta S & \simeq & {\Lambda^2 \over M_P} \int d^4 \theta\ E\, N
\nonumber \\
& \simeq & M_P \int d^4 \theta\ E\, N\ ,
\end{eqnarray}
and the hierarchy is destabilized.  Visible-sector
singlets can destabilize the hierarchy if the visible and invisible
sectors couple, even by nonrenormalizable terms suppressed by $M_P$!

\vspace*{2.0ex}
{\it\noindent 3.3\ Effective Potential}
\vglue 0.2cm

\begin{figure}[t]
\epsfysize=2.8in
\hspace*{1.8in}
\epsffile{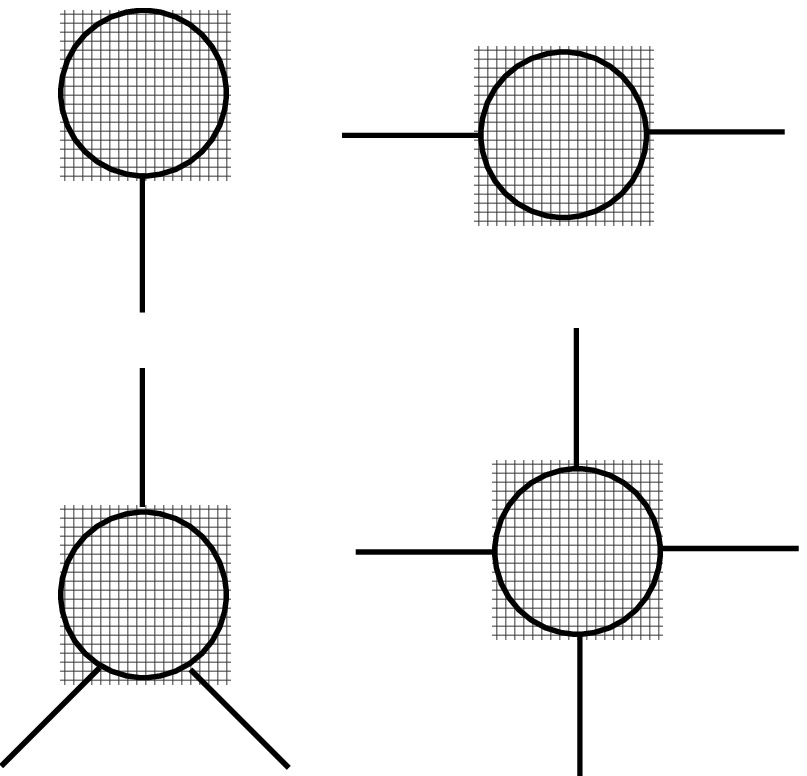}
\fcaption{The superfield effective potential involves vevs
of the hidden-sector fields.}
\end{figure}

Of course, models free from visible-sector singlets are not necessarily
free from destabilizing divergences.  One must still consider the
hidden-sector singlets $C$.  If $\langle C \rangle \lsim M_S +
\theta\theta
M^2_S$, as in dynamical hidden-sector models, the hierarchy is safe.
But if $C$ is a modulus $T$, with a Planck-scale vev, the usual power
counting rules break down, and all graphs are potentially dangerous
(Fig.~4).  In this case one must compute the full effective potential
and make sure that the field-dependent quadratic divergences vanish.

The one-loop correction to the supersymmetric effective potential has
been known for some time \cite{ggrs}.  It is given by
\begin{eqnarray}
\delta S &\simeq & \Lambda^2 \int d^4 \theta \ E\,\log \det K^I{}_J
\ +\ \ldots \nonumber\\
& \simeq & \Lambda^2\ R^I{}_J\, P_I \bar{P}^{J}\ +\ \ldots\ ,
\end{eqnarray}
where $P_I$ is the K\"{a}hler-covariant derivative of the
superpotential, and $R^I{}_J$ is the Ricci tensor of the
K\"{a}hler manifold specified by $K$.  If $R^I{}_J$ does not
vanish, the vacuum can be destabilized,
\begin{eqnarray}
\langle T \rangle & \rightarrow & 0 \nonumber \\
\langle T\rangle & \rightarrow & M_P\ +\ \theta \theta M_P^2 \ .
\end{eqnarray}
In this case the gravitino mass is driven to zero or $M_P$.
Therefore if $R^I{}_J \neq 0$, radiative corrections can lead
to a radically new vacuum.

In a recent paper, Ferrara, Kounnas and Zwirner \cite{lhc} imposed
a geometrical condition on $K$ which ensures that $R^I{}_J =
0$.  The resulting models -- which they called LHC models -- are
automatically free from one-loop quadratic divergences.  One would
like to know whether the one-loop cancellation persists to higher
loops, and if not, to find the conditions that stabilize the
hierarchy.

\vspace*{3.0ex}
{\bf\noindent 4. Destabilizing Divergences at Two Loops}

\vspace*{2.0ex}
{\it\noindent 4.1\ General Formalism}
\vglue 0.2cm

These questions provide the motivation for computing the two-loop
supersymmetric effective potential.  The full calculation is rather
involved, so we will focus on one important piece.  We will show
that there is a two-loop, superpotential-dependent, quadratically
divergent contribution to $V_{\rm eff}$.  This term

\begin{itemize}

\item destabilizes the hierarchy in models with visible-sector singlets,
and

\item illustrates the difficulty of maintaining hierarchy in models
with moduli.
\end{itemize}

\begin{figure}[t]
\epsfysize=1.2in
\hspace*{2.1in}
\epsffile{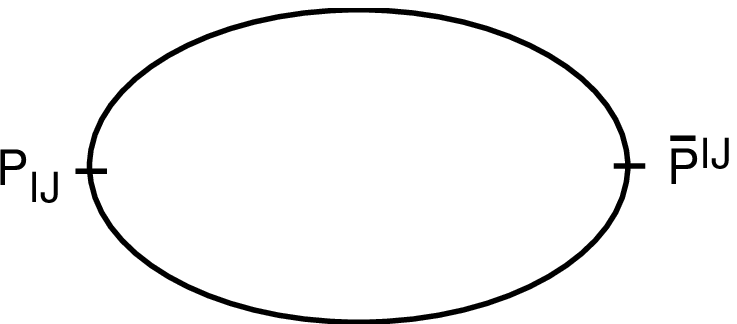}
\fcaption{A one-loop contribution to the superfield effective action.}
\end{figure}

To check our techniques, and to get warmed up, we first compute the
one-loop logarithmically divergent diagram shown in Fig.~5.  We
find
\begin{equation}
\delta S \ \simeq \ {\log \Lambda\over 16 \pi^2}
\ \int d^4 \theta\ e^{2K/3M^2_P}\
\Sigma^+ \Sigma\ P_{IJ}\, \bar{P}^{IJ}.
\end{equation}
This term is super-K\"{a}hler-Weyl invariant, as required.  When
appropriately covariantized, it is also locally supersymmetric.  In
components, it agrees with the result of Gaillard and Jain \cite{gaillard}.
And when $\Sigma^+ \Sigma$ gets a vev, it gives the correct
$\beta$-functions for the soft supersymmetry-breaking masses.

With this experience, we are ready to compute the two-loop
diagram shown in Fig.~6.  Following the steps outlined in
Ref.~\cite{bpr}, we find (See also \cite{choi}.)
\begin{equation}
\delta S \ \simeq \ {1\over 6}\,{\Lambda^2\over (16 \pi^2)^2}
\ \int d^4 \theta\ e^{K/M^2_P}
\  P_{IJK}\, \bar{P}^{IJK},
\label{pijk}
\end{equation}
where $\Lambda$ is a momentum-space cutoff.
As before, this expression is super-K\"{a}hler-Weyl invariant.  It
can be made locally supersymmetric with the help of the supergravity
multiplet.

\vspace*{2.0ex}
{\it\noindent 4.2\ Visible-Sector Tadpoles}
\vglue 0.2cm

Let us use this result to revisit each of the dangerous cases discussed
previously.  For the case of the visible-sector singlet, we can use the
field redefinition
\begin{equation}
H \ \rightarrow\ H \left(1\ -\ {N\over M_P}\right)
\end{equation}
to write (\ref{PN}), (\ref{minimal}) and (\ref{nonminimal}) in the
following form,
\begin{eqnarray}
K & =& N^+ N\ +\ C^+ C\ +\ H^+ H \ +\ {\cal O}(1/M^2_P) \nonumber \\
P & =& {1\over2}\ \mu\, H^2 \ +\ {1\over2}\ m\, N^2\ + \ \lambda^\prime \,
N H^2 \ - \ \lambda^{\prime\prime}\,{N^2 H^2 \over M_P}\ +
\ {\cal O}(1/M^2_P)\ .
\end{eqnarray}
Substituting into (\ref{pijk}), we find the following destabilizing
divergence,
\begin{eqnarray}
\delta S & \simeq & \Lambda^2\ \int d^4 \theta\ e^{K/M^2_P}\ P_{IJK}
\bar{P}^{IJK} \nonumber\\
 & \simeq & \Lambda^2\ \int d^4 \theta\ e^{K/M^2_P}\ {N \over M_P} \ +
\ \ldots \nonumber\\
 & \simeq & {1 \over M_P} \int d^4 \theta\ C^+ C \,N \nonumber \\[2mm]
 & \simeq & M_W\,M^2_S\ n\ .
\end{eqnarray}
This is a two-loop destabilizing divergence.  It indicates that the
one-loop cancellation was purely accidental, and that visible-sector
singlets are always dangerous if they have renormalizable couplings
to the other visible-sector fields.

\begin{figure}[t]
\epsfysize=1.2in
\hspace*{2.1in}
\epsffile{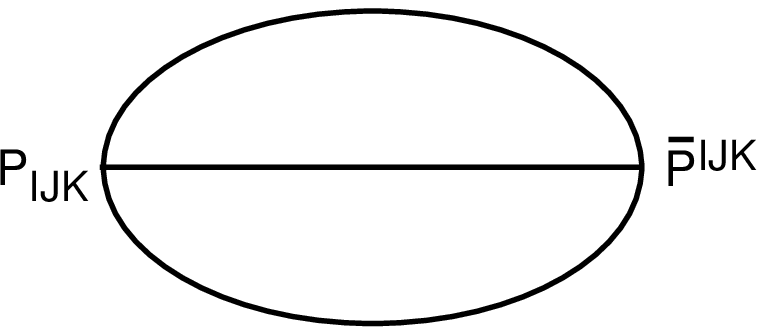}
\fcaption{A two-loop contribution to the superfield effective action.}
\end{figure}

\vspace*{2.0ex}
{\it\noindent 4.3\ Effective Potential}
\vglue 0.2cm

Finally, let us discuss two-loop effective potential.  Equation
(\ref{pijk}) contains a two-loop field-dependent quadratic divergence
-- a divergence which depends on the parameters of the visible-sector
superpotential.  If the hierarchy is to be stable, the divergence
must be canceled, either by contributions from the hidden sector
or by higher string modes.

In either case, a string miracle is required.  The hidden sector
or the higher modes must know about visible-sector parameters like
$m_t,\ m_e,\ V_{cb}$.  These interrelations cannot be understood
in terms of the low energy effective theory.  Our result shows
that string moduli with Planck-scale vevs have the potential to
destabilize the hierarchy.  Whether they do, or not, depends on
the miracles of string theory.

\vspace*{3.0ex}
{\bf\noindent 5. Conclusion}
\vglue 0.2cm

In this talk we considered questions of naturalness in
supersymmetric effective theories.  In particular, we studied
the potentially destabilizing quadratic divergences that are
induced at one- and two-loops in the effective potential.

We first discussed the possible generation of divergent tadpole
diagrams.  We explained why gauge- and global-symmetry singlets
do {\it not} develop tadpoles at one-loop order, if the
K\"{a}hler potential is minimal.  We then showed
that at two loops, quadratically divergent tadpoles can
indeed appear.

Our results indicate that the one-loop cancellation is an accident
of the one-loop approximation.  This conclusion is in accord with
our notions of naturalness because there is no symmetry that
would forbid a divergent tadpole.  Since there is no symmetry,
it has to appear, and indeed it does.

We also discussed the LHC models of Ferrara,
Kounnas and Zwirner.  These models rely on a cancellation of
the one-loop field-dependent quadratic divergences.  Again,
since this cancellation is not related to a symmetry of the
theory, we expect a contribution to arise at higher loops.
Indeed, at two loops we found that the cancellation is spoiled
by terms that depend on the superpotential of the visible sector.

Clearly, if LHC models are to work, there must be some string-induced
conspiracy which cancels the Yukawa-dependent divergence.  Such a
cancellation would be difficult to understand at the level of
effective field theory.

Our conclusions can be readily generalized to higher loops.  The
superpotential-dependent divergences can be guessed by induction.
At one loop, we found logarithmically divergent contributions to
the component K\"{a}hler potential which go like
\begin{equation}
\log \Lambda ~e^{K/M^2_P}\ \left(\,P_{IJ} \bar{P}^{IJ}
{}~+~ {1\over M^2_P}\, P_I \bar{P}^I
{}~+~ {1\over M^4_P}\, P \bar{P}\right)\,\ .
\end{equation}
At two loops, we found quadratically divergent terms such as
\begin{equation}
\Lambda^2~e^{K/M^2_P}\ \left(\,P_{IJK} \bar{P}^{IJK}
{}~+~ {1\over M^2_P}\, P_{IJ} \bar{P}^{IJ}
{}~+~ {1\over M^4_P}\, P_I \bar{P}^I
{}~+~ {1\over M^6_P}\, P \bar{P}\right)\,\ .
\end{equation}
Therefore at three loops, we expect quartically divergent
terms of the form
\begin{equation}
\Lambda^4~e^{K/M^2_P}\ \left(\,P_{IJKL} \bar{P}^{IJKL}
{}~+~ {1\over M^2_P}\, P_{IJK} \bar{P}^{IJK}
{}~+~ {1\over M^4_P}\, P_{IJ} \bar{P}^{IJ}
{}~+~ {1\over M^6_P}\, P_I \bar{P}^I
{}~+~ {1\over M^8_P}\, P \bar{P}\right)\,\ .
\end{equation}
In each case,
the leading term comes from a rigid supersymmetry graph,
while the other terms come from graphs with supergravity
fields in the loops.

Taking $\Lambda \simeq M_P$, we see that the three-loop terms
induce new possibilities for destabilizing divergences.  For
example, the $P_{IJKL} \bar{P}^{IJKL}$ term can also contain
a quadratically divergent tadpole.  This implies
we must expect new superpotential-dependent divergences at
each order of perturbation theory.  The cancellation of quadratic
divergences requires a grand conspiracy between terms at all
orders in the loop expansion.  Presumably this cancellation
is related to the cosmological constant problem, about which
we have nothing to say.

This work was supported by the U.S. National Science Foundation,
grant PHY94-04057.

\newpage

\renewenvironment{thebibliography}[1]{
{\bf\noindent References}
\vglue 0.2cm
\list
 {[\arabic{enumi}]}{\settowidth\labelwidth{[#1]}\leftmargin\labelwidth
 \advance\leftmargin\labelsep
 \usecounter{enumi}}
 \def\newblock{\hskip .11em plus .33em minus .07em}
 \sloppy\clubpenalty4000\widowpenalty4000
 \sfcode`\.=1000\relax}


\begin{thebibliography}{99}

\newcommand{\ib}[3]{ {\em ibid. }{\em #1} (19#2) #3}
\newcommand{\np}[3]{ {\em Nucl.\ Phys. }{\em #1} (19#2) #3}
\newcommand{\pl}[3]{ {\em Phys.\ Lett. }{\em #1} (19#2) #3}
\newcommand{\pr}[3]{ {\em Phys.\ Rev. }{\em #1} (19#2) #3}
\newcommand{\prep}[3]{ {\em Phys.\ Rep. }{\em #1} (19#2) #3}
\newcommand{\prl}[3]{ {\em Phys.\ Rev.\ Lett. }{\em #1} (19#2) #3}
\newcommand{\ap}[3]{ {\em Ann.\ Phys. }{\em #1} (19#2) #3}
\newcommand{\mpl}[3]{ {\em Mod.\ Phys.\ Lett. }{\em #1} (19#2) #3}

\bibitem{gg}L. Girardello and M. Grisaru, \np{B194}{82}{65}.

\bibitem{guts}{J.~Polchinski and L.~Susskind, \pr{D26}{82}{3661};\newline
H.~Nilles, M.~Srednicki and D.~Wyler, \pl{B124}{83}{337};\newline
A.~Lahanas, \pl{B124}{83}{341}.}

\bibitem{old}{U. Ellwanger, \pl{B133}{83}{187};\newline
J. Bagger and E. Poppitz, \prl{71}{93}{2380}.}

\bibitem{jain}V. Jain, \pl{B351}{95}{481}.

\bibitem{bpr}J. Bagger, E. Poppitz and L. Randall, hep-ph/9505244,
to appear in {\em Nucl. Phys.} {\em B}.

\bibitem{ggrs}S.J. Gates, Jr., M. Grisaru, M. Rocek and W. Siegel,
 {\em Superspace}, (Benjamin/Cummings, 1983).

\bibitem{lhc}S. Ferrara, C. Kounnas and F. Zwirner, \np{B429}{94}{589}.

\bibitem{gaillard}M.K. Gaillard and V. Jain, \pr{D46}{94}{1951}.

\bibitem{choi}K. Choi, J. Kim and H. Nilles, \prl{73}{94}{1758}.

\end{thebibliography}
\end{document}